\documentclass[aps,prb,twocolumn]{revtex4-1}
\usepackage[english]{babel}
\usepackage{amsmath,amssymb}
\usepackage{graphicx}
\usepackage{color}

\renewcommand{\H}{\mathfrak{H}}
\renewcommand{\O}{\mathcal{O}}

\newcommand{\Span}{\mathrm{Span}}

\newcommand{\ket}[1]{|#1\rangle}
\newcommand{\bra}[1]{\langle#1|}
\newcommand{\braket}[2]{\langle#1|#2\rangle}
\newcommand{\eps}{\varepsilon}

\begin{document}

\title{Implementation of rigorous renormalization group method for ground space and low-energy states of local Hamiltonians}

\date{\today}

\author{Brenden Roberts}
\email{broberts@caltech.edu}
\author{Thomas Vidick}
\author{Olexei I.~Motrunich}
\affiliation{Institute for Quantum Information and Matter,\\California Institute of Technology, Pasadena, CA 91125}

\begin{abstract}
The success of polynomial-time tensor network methods for computing ground states of certain quantum local Hamiltonians has recently been given a sound theoretical basis by \textcite{arad17}. The convergence proof, however, relies on ``rigorous renormalization group'' (RRG) techniques which differ fundamentally from existing algorithms. We introduce a practical adaptation of the RRG procedure which, while no longer theoretically guaranteed to converge, finds MPS ansatz approximations to the ground spaces and low-lying excited spectra of local Hamiltonians in realistic situations. In contrast to other schemes, RRG does not utilize variational methods on tensor networks. Rather, it operates on subsets of the system Hilbert space by constructing approximations to the global ground space in a tree-like manner. We evaluate the algorithm numerically, finding similar performance to DMRG in the case of a gapped nondegenerate Hamiltonian. Even in challenging situations of criticality, or large ground-state degeneracy, or long-range entanglement, RRG remains able to identify candidate states having large overlap with ground and low-energy eigenstates, outperforming DMRG in some cases.
\end{abstract}

\maketitle

\section{Introduction}
\label{sec:intro} 

Many important techniques for solving lattice models in condensed matter physics take the form of tensor network algorithms. The seminal such method is White's density matrix renormalization group (DMRG),\cite{white92} an optimization algorithm on the matrix product state (MPS) ansatz of quantum wavefunctions\cite{verstraete08,schollwock11} developed as a controllable method improving on Wilson's numerical renormalization group for impurity systems.\cite{wilson75} DMRG remains the most versatile procedure in its class. It has been heavily used to numerically solve low-dimensional quantum models; an early instance was the Haldane phase in the Heisenberg chain.\cite{white94} More recently, a related approach was used in the classification of all gapped phases in one dimension.\cite{chen11} Other related techniques include the tensor renormalization group and tensor network renormalization, which utilize a two-dimensional coarse-graining process to solve quantum systems in one dimension by the quantum-to-classical correspondence.\cite{levin07,gu08,evenbly15} Other variational algorithms operate on the multiscale entanglement renormalization ansatz (MERA), which efficiently represents states exhibiting logarithmic violation of the area law by encoding correlations at all scales in an optimized quantum circuit of logarithmic depth.\cite{vidal07,evenbly09,pfeifer09,evenbly10,evenbly15-1}

In this paper, we present an alternative approach to the solution of local Hamiltonians in one dimension (1D). The rigorous renormalization group (RRG) is a recent algorithm developed as part of a proof of the tractability of computing ground states of gapped local Hamiltonians in 1D.\cite{arad17,landau15} The proof employs techniques first introduced to establish an improved one-dimensional area law.\cite{arad12}
Broadly, the strategy is as follows: partition the system into small initial blocks, and, focusing on the Hilbert space of the blocks individually, identify sets of states that are ``extendable'' to the rest of the system to create a good approximation to the system-wide ground space. This property is termed {\it viability}, and formally defined in Eq.~\eqref{eq:def-viable1}. The identification of viable sets is accomplished with an {\it approximate ground state projector} (AGSP), an operator approximately filtering out highly excited states on the entire system, whose support is restricted to perform this filtering within each block individually. In this way RRG deviates from a traditional real-space blocking scheme, in which each block does not have access to global information. The next step is to merge the identified viable sets on adjacent blocks, obtaining states supported on blocks of larger size. However, this step and the local application of the AGSP result in an untenable blow-up of the number of states, so a reduction step is performed, returning the number of states per block (now comprising two blocks of the smaller size) to a constant value. This procedure is iterated, merging blocks in a tree-like manner, and at the full system scale, the identified states are shown to closely approximate the low-energy space.\cite{arad17}

In the present work we adapt these techniques to specify a concrete RRG procedure allowing for the explicit computation of ground and low-energy states of local Hamiltonians. This requires making allowance for computational limitations, and generally our modifications operate outside of the regime of rigorous guarantee. Still, our algorithm presents a conceptually new approach to this task. We emphasize that the use of the word ``rigorous'' is in reference to the title of \textcite{arad17}, rather than in order to establish a contrast with other tensor network algorithms.

The main conceptual departure of this algorithm from existing tensor network methods is that RRG operates on viable sets of states supported on blocks, rather than on variational states in the full Hilbert space. Two important features arise from this distinction. First, no local energy minimization on a particular ansatz state is performed. Even though in the RRG procedure described here the basic operations are performed on MPS comprising an approximate basis of the viable sets, the MPS objects themselves are incidental, and the concerns arising from the MPS ansatz (e.g., gauge choice, truncation) are external to the fundamental algorithm.

Second, the physical degrees of freedom are not coarse-grained. The objective of a coarse-graining strategy is to limit the dimensionality of the Hilbert space at increasing scale by the introduction of renormalized degrees of freedom, determined by some local rule, specifying a smaller effective Hilbert space. Instead, RRG achieves this goal by maintaining viable sets of constant dimension at all levels of the algorithm hierarchy. These processes cannot be considered equivalent, as the RRG step of applying the AGSP operator changes the relationship between scales in a complicated way, and does not match the intuition of an ``RG flow'' in a small number of parameters. However, this method still allows for fully controllable systematic improvements in accuracy.

The structure of this paper is as follows. We first give a detailed, self-contained description of our algorithm in Sec.~\ref{sec:operation}. (We refer the reader familiar with the theoretical RRG paper to App.~\ref{sec:cspaper} for a precise discussion of the differences between the proof and the present work.) We provide an extended discussion of numerical results in Sec.~\ref{sec:results}. Given its origins as a highly technical theoretical algorithm developed in order to obtain provable guarantees, the RRG method performs surprisingly well, often matching the results of standard DMRG implementations and outperforming them in certain difficult cases exhibiting degenerate ground spaces or highly entangled ground states. Finally, we conclude and give directions for further work in Sec.~\ref{sec:discussion}. 

\section{Operation of algorithm}
\label{sec:operation} 
\begin{figure*}[ht]
The input is a local Hamiltonian $H$ acting on $N$ qubits, specified by an MPO. Let $n$, $s$ and $D$ be input parameters.
\begin{enumerate}
\item Initialize:
\begin{enumerate}
\item Construct approximate ground state projector (AGSP) $K$ from Hamiltonian $H$
\item Partition system into contiguous blocks of length $n$, denoted $J^\lambda_0$, $\lambda = 0,1,\ldots,N/n-1$. Obtain $s$-dimensional low-energy eigenspace $V^\lambda_0$ of block Hamiltonian $H^\lambda_0$ for each $\lambda$.
\end{enumerate}

\item For $m = 0,1,\ldots,\log_2(N/n)-1$, denoting an ``RG step'' or scale factor:
\begin{enumerate}
\item Expand: for $\lambda = 0,1,\ldots,N/(2^m n)-1$:
\begin{enumerate}
\item Extract $D^2$ operators $\{A^\lambda_{m,r}\}_{r=1,\ldots,D^2}$ from the AGSP $K$, acting on subsystem $J^\lambda_m$. Operate on the viable set, taking $V^\lambda_m \to W^\lambda_m \equiv \{ A^\lambda_{m,r} V^\lambda_m \}_r$, where $\mathrm{dim}(W^\lambda_m) \leq sD^2$.
\item Compute the restriction of the block Hamiltonian $H^\lambda_m$ to $W^\lambda_m \subset \H^\lambda_m$.
\end{enumerate}

\item Reduce: For $\lambda = 0,2,\ldots,N/(2^m n)-2$:
\begin{enumerate}
\item (Merge) Obtain the tensor product space \mbox{$W^\lambda_m \otimes W^{\lambda+1}_m \subset \H^{\lambda/2}_{m+1}$}, supported on qubits in $J^{\lambda/2}_{m+1} = J^\lambda_m \cup J^{\lambda+1}_m$. Compute the restriction of $H^{\lambda/2}_{m+1}$ to the tensor product set.
\item Obtain $s$-dimensional low-energy eigenspace of the restriction of $H^{\lambda/2}_{m+1}$ to the tensor product space. Use the eigenstates as a basis for viable set $V^{\lambda/2}_{m+1}$ in iteration $m+1$.
\end{enumerate}
\end{enumerate}

\item At $m = m^\ast = \log(N/n)$, the viable set $V^0_{m^\ast}$ is a candidate for the low-energy space $T$ supported on the full system.
\end{enumerate}
\caption{\label{fig:outline}Outline of the implemented RRG algorithm.}
\end{figure*}

\subsection{Overview and notation}

\begin{figure*}[ht]
\includegraphics[width=\textwidth]{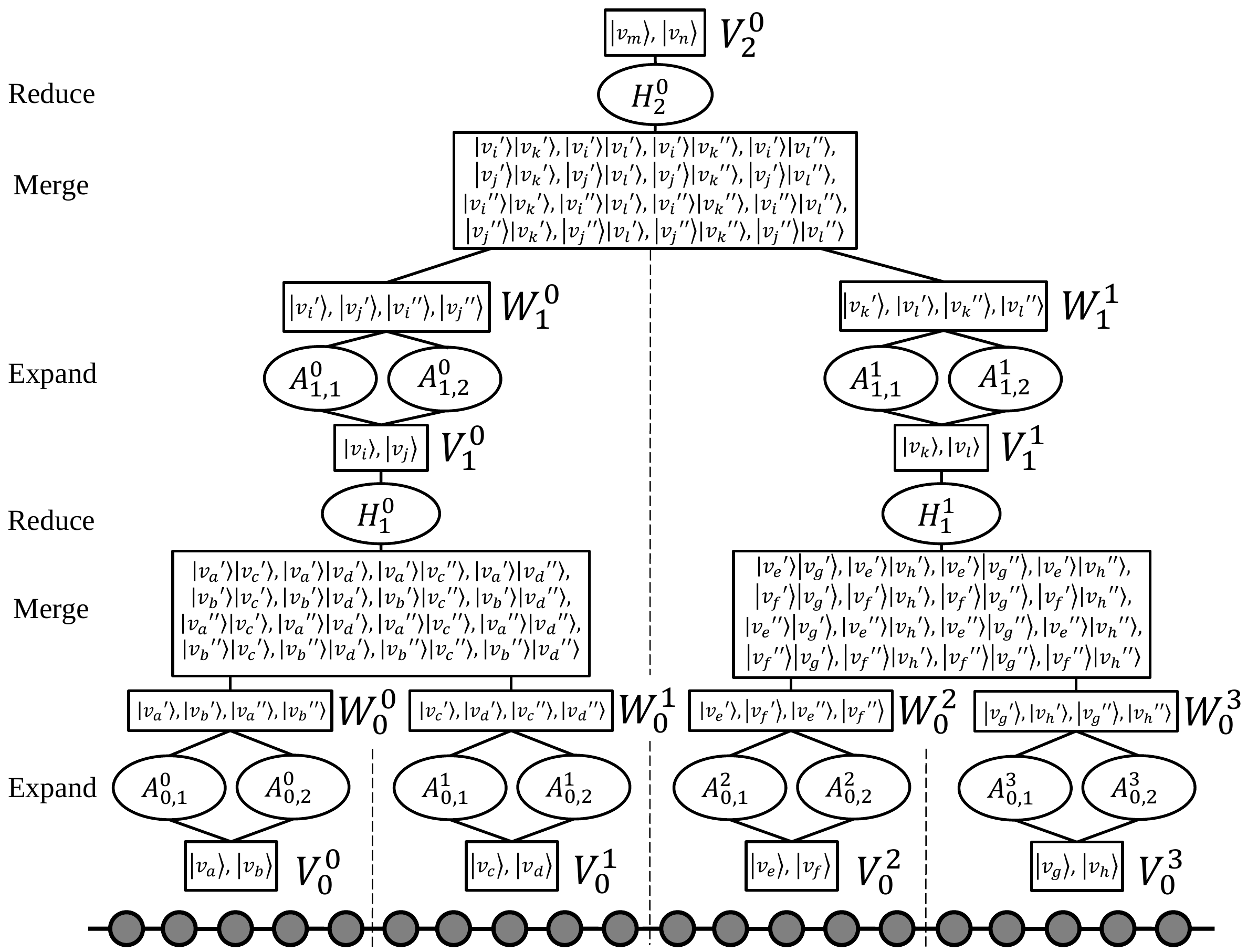}
\caption{\label{fig:schematic}Schematic illustration of the RRG algorithm over several length scales $m=0,1,2$. As shown, $(s,D)=(2,2)$ and block size $n=5$. Gray dots represent local Hilbert spaces, and the $V^0_0$, $V^1_0$, etc, are the viable sets over blocks of sites. The labeled vectors $\ket{v_a}$, etc., are basis states for the viable sets and have no relationship between blocks at a given scale. The action of the projector operators $A^\lambda_{m,r}$ on the states is represented by primes and double primes (e.g., $A^0_{0,1}\ket{v_a} = \ket{{v_a}'}$ and $A^0_{0,2}\ket{v_a} = \ket{{v_a}''}$). These generate the expanded viable sets $W^0_0$, and so on. The Merge procedure obtains tensor product states such as $\ket{{v_a}'}\ket{{v_c}'}$ supported on two blocks, and the tensor product set is reduced in dimension via diagonalizing block Hamiltonians like $H^0_1$, producing a viable set supported on two of the previous blocks.}
\end{figure*}

The steps of RRG as implemented are listed in Fig.~\ref{fig:outline} for reference and are discussed in detail in subsequent sections. A visual schematic is shown in Fig.~\ref{fig:schematic}. Our notation is as follows. Let $H = \sum_{i=0}^{N-2} h_i$ be a 2-local Hamiltonian on a chain of $N$ qubits, with term $h_i$ acting on sites $i$ and $i+1$. (The generalization to $k$-local Hamiltonians and qudits is straightforward.) Denote the Hilbert space of the system by $\H$, and refer to the low-energy eigenspace of $H$ as $T$. Let $n$ be a parameter specifying the size of initial regions of the system, and assume $N/n$ is a power of $2$. For each $m=0,1,\ldots,\log_2 (N/n)$, partition the $N$-site system into contiguous blocks of equal length $2^m n$. Call these \mbox{$J^\lambda_m = \{ \lambda 2^m n , \ldots , (\lambda+1) 2^m n-1 \}$}, for \mbox{$\lambda = 0,1,\ldots,N/(2^m n)-1$}. The Hilbert space associated with $J_m^\lambda$ is denoted $\H^\lambda_m$, and $\H = \bigotimes_\lambda \H^\lambda_m$. Let $H^\lambda_m$ be the block Hamiltonian on $J^\lambda_m$, comprising all terms acting only on sites in $J^\lambda_m$ and excluding boundary terms. Explicitly, $H^\lambda_m = \sum_{i \in {J^\lambda_m}^\ast} h_i$, where \mbox{${J^\lambda_m}^\ast = \{ \lambda 2^m n , \ldots , (\lambda+1) 2^m n -2 \}$}.

\subsection{Initialization}
\label{sec:init}

The first step is to construct an {\it approximate ground state projector} (AGSP) $K$, whose action on states in $\H$ increases overlap with $T$, the low-energy subspace of $H$. Many constructions of AGSP are possible. In the interest of efficiency, we use an AGSP obtained as an approximation to a thermal operator at temperature $t/k$, $K\approx e^{-kH/t}$, $t,k > 0$. Let $Q_t$ denote a matrix product operator (MPO) approximating the thermal operator $e^{-H/t}$ at temperature $t$; procedures such as a Trotter decomposition\cite{suzuki76} or cluster expansion can be used to efficiently compute this MPO. The AGSP is then obtained as a power of $Q_t$, contracting the product on the physical indices $k$ times. 

Because the AGSP must later be divided into operators acting on individual blocks, to compute $Q_t$ requires contraction of the tensor network having terms of the form $e^{-h_i/t}$. After each contraction an SVD is performed between site indices, and the MPO is truncated by eliminating low-weight Schmidt vectors across each bond. Here truncation is meant in the sense of MPS truncation, representing the MPO as a state in a higher-dimensional local Hilbert space. This amounts to using the Frobenius norm to order the terms arising from the SVD, and may not be an optimal way to approximate operators; we address this issue in more detail later.

The second step in the initialization is to identify sets of states $V^\lambda_0 \subset \H^\lambda_0$, for $\lambda = 0,1,\ldots,N/n-1$, of constant dimension $s$, where $s$ is a parameter of the algorithm which bounds the dimension of the sets manipulated throughout. We use the term ``viable sets'' for the $V^\lambda_0$ (generally, for $V^\lambda_m \subset \H^\lambda_m$)  because the intent of the algorithm is that each $V^\lambda_m$ be extendable to include a good approximation to the global low-energy eigenspace $T$. That is, each set $V^\lambda_m$ is chosen such that if $\H = \H^\lambda_m \otimes \overline \H^\lambda_m$, then $V^\lambda_m \otimes \overline \H^\lambda_m$ contains a subspace which is a good approximation to $T$. We identify a set $V^\lambda_m$ as {\it $\delta$-viable} if
\begin{equation}\label{eq:def-viable1}
P_T P_{V^\lambda_m \otimes \overline \H^\lambda_m} P_T \geq (1 - \delta) P_T,
\end{equation}
where $P_T$ is a projector onto a subspace $T$. More concretely, consider the case of a non-degenerate global ground space \mbox{$T = \Span\{\ket \tau \}, \ket \tau \in \H$.} The viability of the set $V^\lambda_m$ is given by
\begin{equation}\label{eq:def-viable2}
\delta =  1 - \max_{\ket x \in V^\lambda_m \otimes \overline\H^\lambda_m} |\braket{\tau}{x}|^2\;,
\end{equation}
where $\ket x = \sum_j a_j \ket{v_j}\ket{\overline v_j}$ for a collection of states $\{\ket{v_j}\} \subset V^\lambda_m$ along with coefficients $a_j$, and states $\{ \ket{\overline v_j} \}$ arbitrary in the Hilbert space of the sites in the complement. It will be shown in Sec.~\ref{sec:iter} that one need never explicitly compute the $\{ \ket{\overline v_j} \}$. For the case that $\mathrm{dim}(T) > 1$, $\delta$ is obtained by taking the smallest value of the maximum in~\eqref{eq:def-viable2}, over all \mbox{$\ket{\tau} \in T$.} The goal of the algorithm is to construct the viable sets $V^\lambda_m$ in such a way that they are indeed $\delta$-viable for some constant $\delta$ less than 1 for all scales $m$. Note that a small value of $\delta$ corresponds to a better approximation, in contrast with measures like overlap. We emphasize that the viability parameter is not explicitly computed by the algorithm. Instead, it provides a useful metric to evaluate performance, both in terms of the theoretical results and in terms of experimental investigations for cases where we wish to compare with other methods providing estimates for the ground space (such as when exact diagonalization is possible).

If $n$ is chosen to be small enough, generic operators on $\H^\lambda_0$ can be exactly diagonalized. In the initialization step, the initial viable set $V^\lambda_0$ is specified to be the span of the $s$ eigenvectors of $H^\lambda_0$ of lowest energy, obtained by exact diagonalization.

\subsection{Iteration over scale}
\label{sec:iter}

The algorithm proceeds through a tree-like hierarchy, the levels of which are specified by a scale parameter \mbox{$m = 0,1,\ldots,\log_2(N/n)$}. At scale $m$, block $J^\lambda_m$ consists of $2^m n$ sites and the region index $\lambda$ runs from $0$ to $N/(2^m n)-1$. Note that although the scale of the algorithm is increasing, we do not eliminate any of the physical degrees of freedom. At each step we assume that the previous level has produced a viable set $V^\lambda_m$ with basis $\{\ket{v_q}\}_{q=1,\ldots,s}$ represented by MPS, for every $\lambda$.

The algorithm performs two steps. The first step is the {\it expansion} of the viable set, which has the effect of improving the viability parameter $\delta$ as defined in Eq.~\eqref{eq:def-viable1}. This is accomplished using the AGSP constructed in the initialization step as follows. Let $J^\lambda_{m,L}$ denote the qubits to the left of $J^\lambda_m$, and $J^\lambda_{m,R}$ those to the right. (Generally $J^\lambda_m$ has two boundaries with its complement $J^\lambda_{m,L} \cup J^\lambda_{m,R}$. The system-edge cases follow immediately.) Consider the MPO representation of the AGSP $K$, whose elementary tensors are collections of operators on the local Hilbert space, as an MPS. The Schmidt decomposition of $K$ across the left boundary, separating $J^\lambda_{m,L}$ from $J^\lambda_m \cup J^\lambda_{m,R}$, produces a virtual index of dimension $\zeta$:
\begin{equation}\label{eq:K1}
K = \sum_{\alpha < \zeta} \sigma_\alpha L_\alpha M_\alpha.
\end{equation}
The $L_\alpha$ are the left Schmidt vectors and the $M_\alpha$ the right---which are operators on \mbox{$J^\lambda_m \cup J^\lambda_{m,R}$}---each with a corresponding Schmidt coefficient $\sigma_\alpha$. The Schmidt decomposition may then be obtained for each of the $M_i$ across the boundary between $J^\lambda_m$ and $J^\lambda_{m,R}$, producing a virtual index of dimension $\xi$. That is,
\begin{equation}\label{eq:K2}
M_\alpha = \sum_{\beta < \xi} \nu_{\alpha\beta} A_{\alpha\beta} R_{\alpha\beta}
\end{equation}
Each $A_{\alpha\beta}$ is an operator on $\H^\lambda_m$, with weight \mbox{$\gamma_{\alpha\beta} = \sigma_\alpha \nu_{\alpha\beta}$} in the expansion of $K$. For clarity we make the algorithm variables explicit: $A^\lambda_{m,\alpha\beta}$. Now let $D>0$ be another parameter of the algorithm. In order to increase the viability of the set $V^\lambda_m$, act on it with the $D^2$ operators \mbox{$A^\lambda_{m,r}$, $r = (\alpha,\beta)$}, having highest weight $\gamma_r = \gamma_{\alpha\beta}$. That is, take \mbox{$V^\lambda_m \to W^\lambda_m = \Span( \{A^\lambda_{m,r} \ket{v_q} \}_{r,q})$}, which we refer to as an {\it expanded viable set} with dimension bounded by $sD^2$.

One expects this operation to produce a set $W^\lambda_m$ of better viability than $V^\lambda_m$ because the $A^\lambda_{m,r}$ operators together are meant to increase overlap with the global low-energy space $T$: this is the defining property of the AGSP. More precisely, let $\{\ket{v_j}\}$ be a collection of states in $V^\lambda_m$ such that there exists $\{\ket{\overline v_j}\} \in \overline \H^\lambda_m$ such that for some coefficients $a_j$, the state $\ket{x} = \sum_j a_j \ket{v_j} \ket{\overline v_j}$ has good overlap with $T$. By construction, $K \ket{x}$ has better overlap with $T$ than $\ket x$. Using the decomposition of Eqs.~\eqref{eq:K1} and \eqref{eq:K2},
\begin{equation}\label{eq:Kx}
K\ket{x} = \sum_{\alpha,\beta,j} \gamma^\lambda_{m,\alpha\beta} a_j A^\lambda_{m,\alpha\beta} \ket{v_j} \otimes \overline A^\lambda_{m,\alpha\beta}\ket{\overline v_j},
\end{equation}
where $\overline A^\lambda_{m,\alpha\beta} = L^\lambda_{m,\alpha} R^\lambda_{m,\alpha\beta}$. In this way the viability as defined in Eq.~\eqref{eq:def-viable2} of the set $V^\lambda_m$ can be improved while leaving both the states and the operators supported on the complement $\overline \H^\lambda_m$ entirely implicit.

If all operators $A^\lambda_{m,\alpha\beta}$ were applied to $V^\lambda_m$, the resulting set would contain the collection of states $\{A^\lambda_{m,\alpha\beta} \ket{v_j}\}$, which has improved viability. However, instead of applying all $A^\lambda_{m,\alpha\beta}$, which would lead to an unmanageable blow-up in the size of the viable set, we introduce an approximation by selecting the $D^2$ operators $A^\lambda_{m,r}$ of highest weight $\gamma_r$ in order to obtain $W^\lambda_m$. There is no formal guarantee that this is the best choice, as the Schmidt decomposition is based on the Frobenius rather than the operator norm. In practice we found the choice to be quite reasonable: to observe the increase in viability in a nondegenerate gapped model, compare the $V$ and $W$ points in Fig.~\ref{fig:nis_prog}, and in a critical model in Figs.~\ref{fig:tfim_crit}, \ref{fig:tfim_prog}.

The second step performed at each scale $m$ is that of {\it reduction} of the dimension of the expanded viable sets $W^\lambda_m$ and $W^{\lambda+1}_m$ to generate $V^{\lambda/2}_{m+1}$. At the cost of a loss of viability, this step restores $s$-dimensionality, resulting in a viable set suitable to use at the next level. One first performs a {\it merge} operation on disjoint pairs of blocks $(\lambda, \lambda+1)$, with $\lambda =  0,2,\ldots,N/(2^m n)-2$. Merging refers to computing the tensor product set $W^\lambda_m \otimes W^{\lambda+1}_m$ that has support on sites $J^\lambda_m \cup J^{\lambda+1}_m$. One obtains the viable set $V^{\lambda/2}_{m+1}$, a subspace of $\H^{\lambda/2}_{m+1} = \H^\lambda_m \otimes \H^{\lambda+1}_m$, from the $s$-dimensional low-energy eigenspace of the restriction of $H^{\lambda/2}_{m+1}$ to $W^\lambda_m \otimes W^{\lambda+1}_m$. We note that this step differs from its counterpart in the theoretical algorithm, which proceeds via random sampling instead of deterministically selecting the lowest-energy eigenvectors of $H^{\lambda/2}_{m+1}$, as we do here. Our choice is based on efficiency considerations described below; see also Appendix \ref{sec:cspaper} for further discussion. The effect of the operation on the viability of the reduced subspaces can be seen in Figs.~\ref{fig:nis_prog}, \ref{fig:tfim_crit}, and \ref{fig:tfim_prog}.

The single viable set $V^0_{m^\ast}$ generated at $m^\ast = \log_2 (N/n)$ after the reduction step at scale $m^\ast-1$, is a constant-dimensional $\delta$-viable subspace with support on the full system. The algorithm returns the $s$ lowest-energy eigenvectors of the restriction of $H$ to $W^0_{m^\ast-1} \otimes W^1_{m^\ast -1}$, which comprise a basis for this candidate subspace.

\subsection{Scaling and computational considerations}

The accuracy with which RRG approximates low-energy eigenstates of $H$ is controlled primarily by two parameters, $s$ and $D$. To recapitulate, $s$ bounds the dimension of the reduced viable sets at each step, and $D$ controls the level of approximation in the application of the AGSP via the operators $\{A^\lambda_{m,r}\}$, $r=1,\ldots,D^2$. Both parameters are reflected in the bound on the dimension $sD^2$ of the expanded viable sets $W^\lambda_m$.

We review the steps in the algorithm and discuss their complexity scaling based on these parameters. In addition to $s$ and $D$, important parameters are the system size $N$ and the bond dimensions $\chi$ for MPS and $\eta$ for MPO that are manipulated throughout. For physical Hamiltonians it is reasonable to expect $\chi$ and $\eta$ to be constant in the gapped case, and in gapless systems $\chi,\eta \sim N$. See~\textcite{schollwock11} for a discussion of the scaling of basic MPS operations. Note that the initial block size $n$ only enters this analysis in determining the number of necessary layers $\log(N/n)$.

The initialization requires obtaining viable sets $V^\lambda_0$ of the Hilbert space $\H^\lambda_0$ on the qubits $J^\lambda_0$. For small enough choices of $n$ the complexity of this step will be negligible, so we omit it. Similarly, the computation of the full AGSP \mbox{$K \approx (e^{-H/t})^k$} can be done efficiently via Trotter decomposition, and is not an important bottleneck. In order to extract the operators $\{A^\lambda_{m,r}\}$, $r=1,\ldots,D^2$, the AGSP must be obtained as an MPO in canonical form, analogous to that used for MPS. To do so requires a sequence of $\O(N)$ SVD operations, each with cost $\O(\eta^3)$. 

For the steps comprising the iterated procedure we give scaling results applicable at the final computational level $m = m^\ast-1$. The first step is to apply $K$ to each $V^\lambda_m$ by means of the Schmidt decomposition of $K$ across the boundary separating $J^\lambda_m$ from its complement $\bigcup_{\lambda' \neq \lambda} J^{\lambda'}_m$. This yields a set of operators acting on $\H^\lambda_m$. Applying the $D^2$ such operators of highest Schmidt weight to a basis of the subspace takes $V^\lambda_m \to W^\lambda_m$, increasing the dimension to $sD^2$. The total cost of contracting these MPS and MPO is $\O(s D^2 N \chi^2 \eta^2)$.

The second step acts on disjoint pairs of neighboring regions, forming the tensor product of expanded viable sets: $W^\lambda_m \otimes W^{\lambda+1}_m$, with dimension $(sD^2)^2$. We compute the matrix elements of the restriction of the block Hamiltonian to the tensor product set. The scaling of this step is $\O\left([(sD^2)^2]^2 N \chi^3\right)$. For local Hamiltonians the constant can be improved using the decomposition
\begin{align}
H^{\lambda/2}_{m+1} &= H^\lambda_m + H^{\lambda+1}_m + B^{\lambda/2}_{m+1}\nonumber\\
&= H^\lambda_m + H^{\lambda+1}_m + \sum_p B^\lambda_{m,p} \otimes B^{\lambda+1}_{m,p}\;.
\end{align}
The operator $B^{\lambda/2}_{m+1}$ contains $\O(1)$ terms in $H$ acting across the boundary between $J^\lambda_m$ and $J^{\lambda+1}_m$.

Exact diagonalization of the restricted block Hamiltonian in the subspace has complexity \mbox{$\O([(sD^2)^2]^3)=\O(s^6 D^{12})$}. After this, the final step is to explicitly compute the $s$ lowest-energy eigenstates, which has a total cost $\O(s (sD^2)^2 N \chi^3)$. These states are used as a basis for the viable set at the next iteration.

From this coarse analysis it is clear that the limiting step with respect to $s$ and $D$ is the diagonalization of the restricted block Hamiltonian. This step is not part of the original formulation, which specifies instead that the reduction of viable set dimension take place by randomly selecting states from the tensor product set. The choice of our variant is motivated by its effect on the entanglement of the intermediate basis states: low-energy excited states of a block Hamiltonian may display lower entanglement than states chosen randomly. In practice this lowers $\chi$ in some systems. It also demonstrates a different possible interpretation of the parameter $s$, which during the iteration step implicitly defines an energy scale with respect to the restricted Hamiltonian. States having block excitation energy higher than this scale are inaccessible to the algorithm for the purposes of the expansion step. 

\section{Numerical results}
\label{sec:results} 

We now present results from RRG for some example models with the following goals in mind. We first validate the algorithm in a simple gapped nondegenerate system in Sec.~\ref{sec:n1}, demonstrating consistency with DMRG as well as previous numerical and perturbation theory results. In this case the states obtained by RRG are of similar accuracy to those of DMRG, with run times a factor of 5--10 slower depending on $s$, $D$,  and $N$. However, we emphasize that it is not the objective of RRG to obtain a numerically precise ground state; rather, it is to accurately identify states having constant overlap with the global low-energy subspace. One expects an optimization algorithm to obtain a more precise state in the absence of local energy minima or very flat energy landscapes, and for simple models we take the DMRG ground state to be exact (in particular, using it to measure viability $\delta$). The RRG candidate states may later be variationally optimized in order to achieve a particular accuracy, but we do not modify the states here.

Our next goal is to demonstrate the practical scaling of the algorithm's performance and computational costs associated with the subspace parameters $(s,D)$. We use the familiar case of the Ising model in the transverse field in Sec.~\ref{sec:n2}, both away from and at criticality. We find that for low values of these parameters, often surprisingly good results can be obtained, with close to unity overlap between DMRG and RRG ground state candidates. However, neither algorithm scales linearly with system size in the critical regime. Here the slowdown of RRG is no longer a simple numerical factor but becomes a significant cost at larger system sizes (beyond a few hundred sites in our implementation) or for larger values of the algorithm parameters. 

Finally, we consider somewhat more challenging models demonstrating areas in which RRG may hold an advantage. In Sec.~\ref{sec:n3} we investigate the Bravyi-Gosset model,\cite{bravyi15} which has $\O(N)$ ground state degeneracy, by obtaining a complete basis for the ground space. In Sec.~\ref{sec:n4} we consider the XY model with randomly-distributed couplings. The ground state of this model, the random singlet phase, displays long-range entanglement in that it supports algebraic decay of correlations. We compare the correlations present in the candidate states of DMRG and RRG to exact results obtained by the Jordan-Wigner transformation, finding that RRG more accurately reproduces observables measured on the state.  

All numerical results were obtained using the tensor network library ITensor\cite{itensor} for both the DMRG and RRG computations. In all of the following, a Trotter decomposition with 60 steps was used to obtain the tensor network for $Q_t \approx e^{-H/t}$, with $t=10$, and degree $k=8$ used to compute the AGSP $K \approx (Q_t)^k$. Thus the effective temperature $t/k$ is of order unity. For reasonable choices of parameters the accuracy of the approximation $Q_t$ is not a limiting factor of the algorithm. Computations were performed on standard hardware on a single node of a computing cluster, with only single threading for the reported run times. A single error parameter $\tau$ was used to control MPS truncation in ITensor for both DMRG and RRG (usually $\tau \sim 10^{-9}$--$10^{-12}$); in most cases a more lenient value would drastically improve run times with little effect on accuracy. DMRG convergence was handled using a fixed number of sweeps $\geq 20$ and relying on the internal diagonalization routine included in ITensor without any modifications specific to the individual systems. Excited states were found iteratively in DMRG by adding projectors into previously-found states to the Hamitonian and using random trial wavefunctions. Often the average viability will be used as a metric; this is simply the average over region label $\lambda$ of the viability $\delta$ of each viable set ($V^\lambda_m$ or $W^\lambda_m$) at fixed level $m$.

\subsection{Nonintegrable Ising model}
\label{sec:n1}
This model refers to a spin-1/2 Hamiltonian
\begin{equation}
H = -J \sum_{i=0}^{N-2} \sigma^z_i \sigma^z_{i+1} - g \sum_{i=0}^{N-1} \sigma^x_i - h \sum_{i=0}^{N-1} \sigma^z_i .
\end{equation}
For $h \neq 0$ the model is gapped with a nondegenerate ground state, and admits no good quantum numbers due to the longitudinal component of the field. A recent numerical study\cite{lin17} for the parameters $(J,g,h) = (1,-1.05,0.5)$ found the ground state energy density to be $\eps_0/N \approx -1.722$ and the gap $\gamma = 3.6401$.

\begin{figure}[h]
\includegraphics[width=\columnwidth]{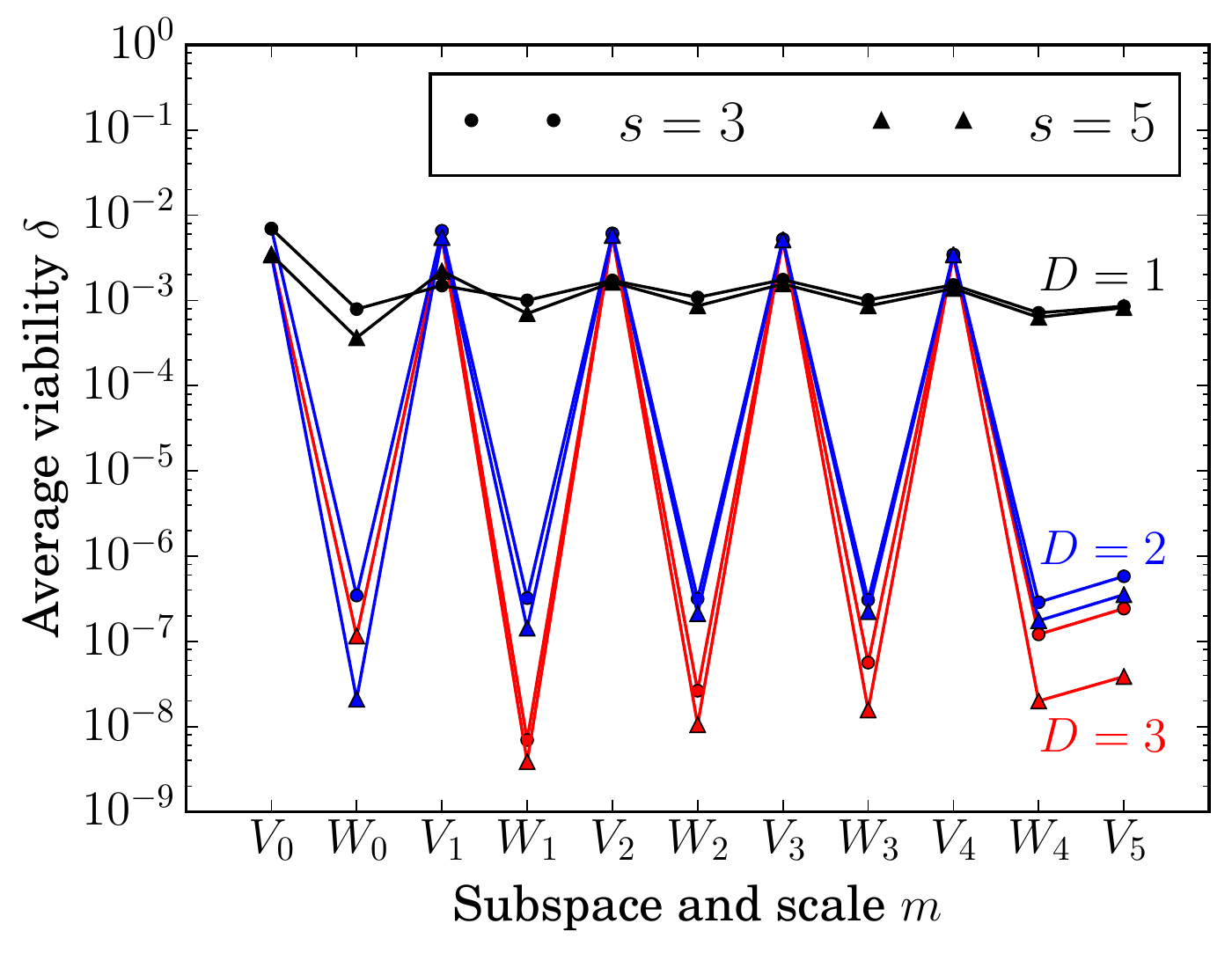}
\caption{\label{fig:nis_prog}(color online) Viability of sets $V^\lambda_m$, $W^\lambda_m$ averaged over $\lambda$, for nonintegrable Ising model with $N=256$ spins, obtained as the RRG algorithm progresses through the scale hierarchy. Data are shown for parameter values $s=3,5$ and $D=1,2,3$.}
\end{figure}

\begin{figure}[ht]
\includegraphics[width=\columnwidth]{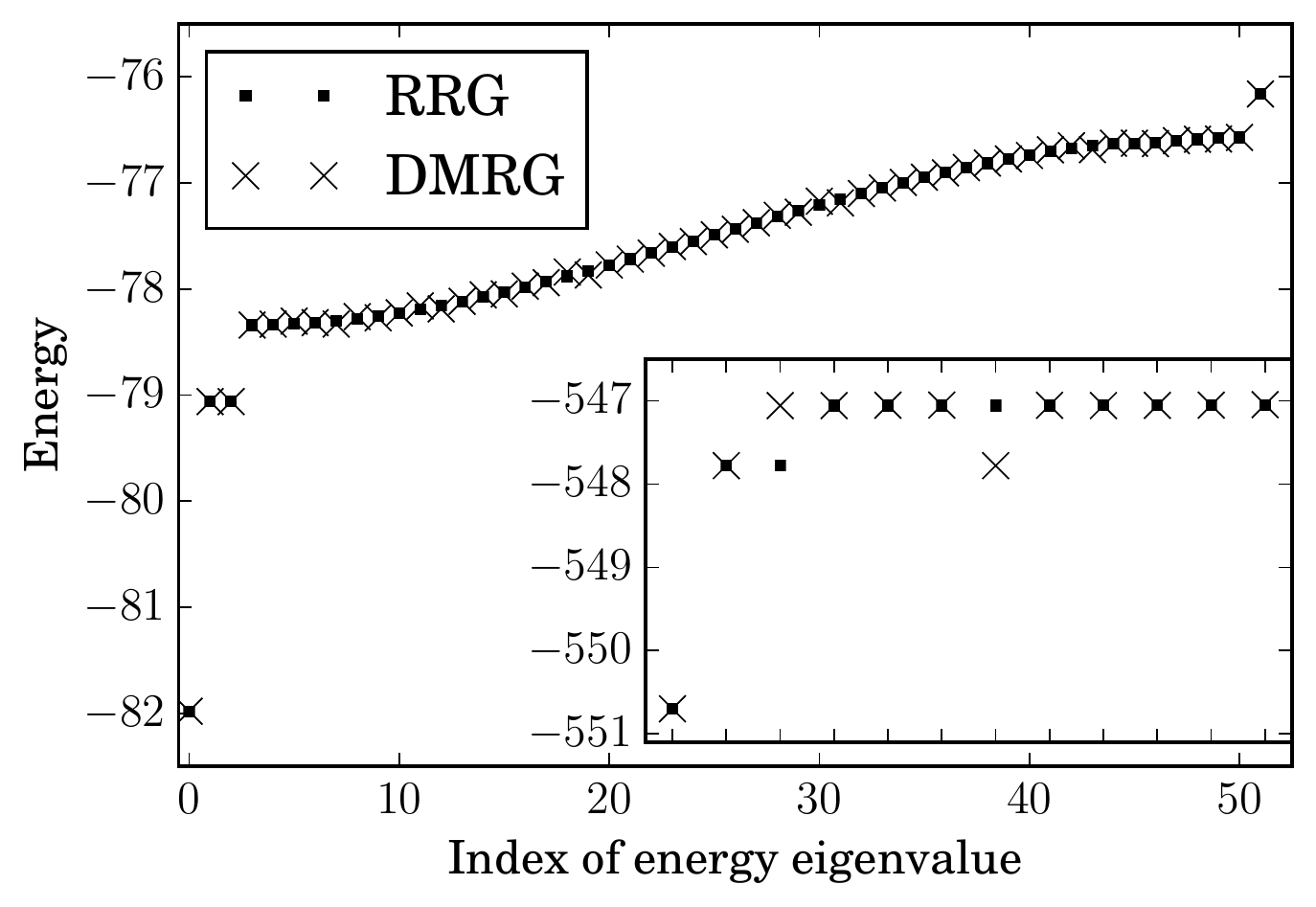}
\caption{\label{fig:nis_spec}Energy eigenvalues of the nonintegrable Ising Hamiltonian for $N=48$ within the subspace obtained by RRG for $(s,D)=(52,3)$, along with DMRG results for low-energy states. Inset: the same computation for $N=320$ and $(s,D)=(12,3)$. DMRG does not consistently identify both edge states in sequence; see text for details.}
\end{figure}

We run the RRG algorithm for a fixed system size $N=256$, initial block size $n=8$, and track the average viability $\delta$ of the viable sets $V_m$ and $W_m$ through the sequence of dimensional expansion and reduction at each scale $m$ (see Fig.~\ref{fig:schematic}). Each data point shown in Fig.~\ref{fig:nis_prog} is the average over $\lambda$ at a given length scale $m$. The parameters $(s,D)$ are varied to demonstrate their influence on the results. For gapped systems of this size both DMRG and RRG have run times scaling linearly with system size, however RRG runs more slowly by a factor of 5--10 compared to DMRG. At $N=256$, DMRG took 5 minutes to converge $s=5$ states (ground and four excited) and RRG ran in 30 minutes with $(s,D)=(5,3)$.

The large improvement in viability from $V$ to $W$ is attributable to the AGSP, rather than simple increase in dimension. Both $\dim(V)=s$ and $\dim(W)=sD^2$ are constant in $m$ and very small compared to the dimensions of the block Hilbert spaces. Choosing $n$ vectors without bias from an $M$-dimensional space will produce a subspace whose squared overlap with a specific vector is of order $n/M$. Since $M$ here is exponentially large, a constant increase in $n$ would not much affect measured viability. Thus, the AGSP is an effective projector even at low values of $D$, which we expect as the model is gapped.

A consequence is that the accuracy of RRG for the largest $(s,D)$ is comparable to that of DMRG, but we do not expect this to be a general feature. Recall from Sec.~\ref{sec:init} that the criterion the algorithm seeks to maintain is that the measured viability $\delta$ of the $V^\lambda_m$ (and thus the average viability) be bounded for all $m$ by some constant $\delta^\ast < 1$, rather than approaching unity exponentially in $m$. The viability of the $W^\lambda_m$ is not necessarily specified, but should be sufficiently good for the $V_{m+1}$ viable sets at the next level to satisfy the bound. For assessing the performance of RRG, as in Fig.~\ref{fig:nis_prog}, one seeks that $\delta$ be maintained away from 1 for the $V_m$ averages.

The final $s$-dimensional viable sets $V_{m^\ast}$ ($V_5$ in Fig.~\ref{fig:nis_prog}) here and in the following examples display much better average viability than that of the previous $V_m$. This is generally true: at steps $m < m^\ast$ the viable set is found by diagonalizing a block Hamiltonian $H^\lambda_m$, which omits terms present in $H$. The low-energy eigenspace of this operator need not be close to $T$, the global low-energy space. At $m = m^\ast$, however, the low-energy eigenspace of $H^0_{m^\ast} = H$ coincides with $T$, resulting in minimal loss of viability from the dimensional reduction.

By changing the parameters of RRG, we obtain candidates for low-energy excited states. The ground state of this model is close to a uniform spin-up state, and the excited band contains a spin-flip excitation. Under open boundary conditions two nearly degenerate lower-energy states separate from the first band, corresponding to quasiparticles localized at either edge. We obtain the low-energy spectrum for $N=48$ with $(s,D) = (52,3)$, and for $N=320$ with $(s,D)=(12,3)$. The results are shown in Fig.~\ref{fig:nis_spec}, compared with DMRG states. For small $N$ both methods find the entire first excited band. In the larger system, the localized edge states are more difficult for DMRG, and it does not consistently find the edge states in sequence. The RRG ground state energy density at $N=320$ is $\eps_0/N = -1.721$ and the gap to the excited band is $\gamma = 3.6402$, in agreement with previous results. We find the half-chain entanglement entropy of the ground state and edge states to be \mbox{$S = 0.01$ bits,} and of the states in the band to be \mbox{$S \approx 1.01$ bits,} consistent with qualitative understanding of these states. For DMRG and RRG, ground states have bond dimension 4 and excited states in the band have bond dimension 31. (The methods do not yield identical bond dimension in all cases.)

\subsection{Transverse-field Ising model}
\label{sec:n2}
\begin{figure}
\includegraphics[width=\columnwidth]{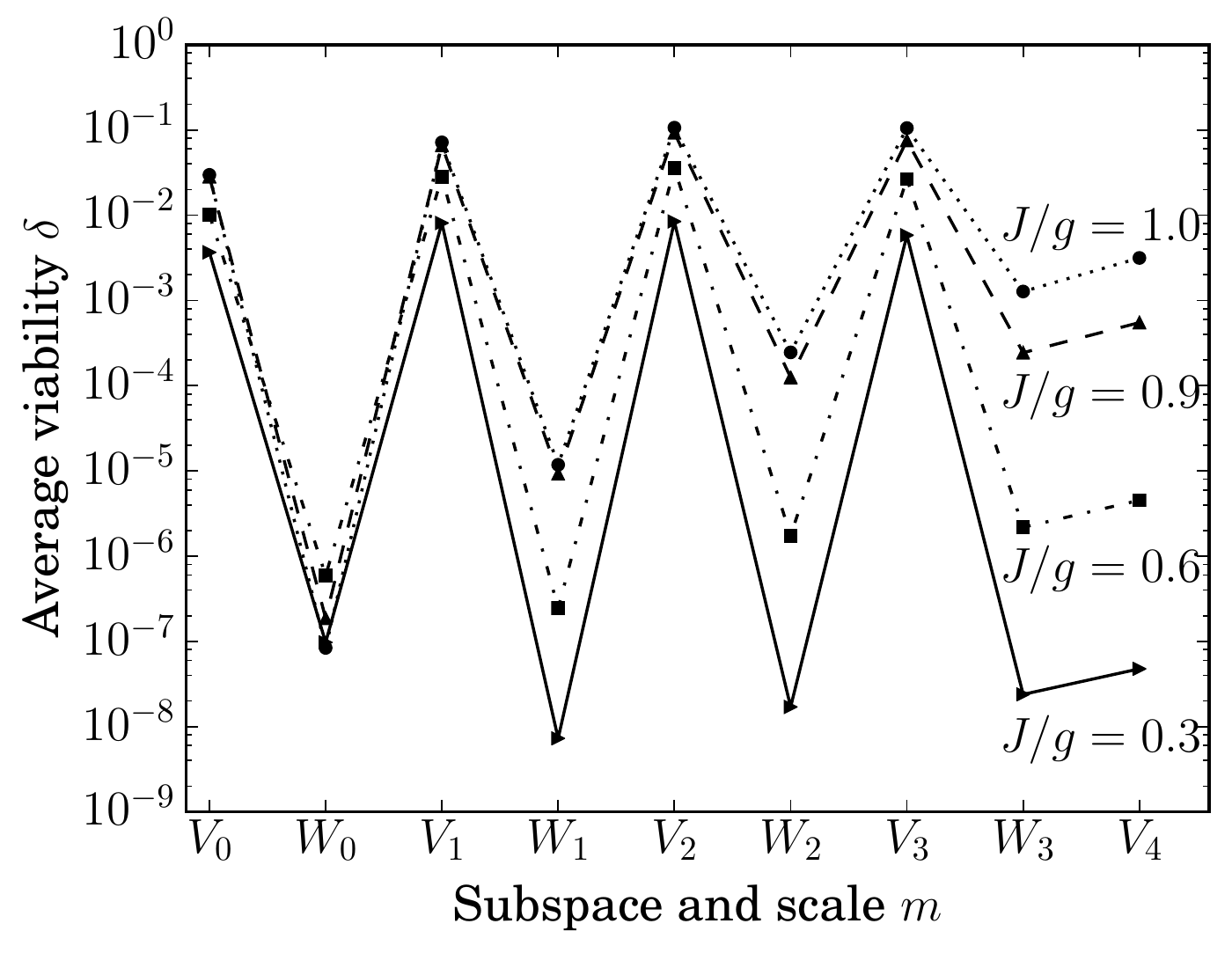}
\caption{\label{fig:tfim_crit}Viability of sets $V^\lambda_m$, $W^\lambda_m$, averaged over $\lambda$, for the transverse-field Ising model both away from and at criticality. The number of spins is $N=128$. All data points were generated using parameter values $(s,D)=(5,4)$.}
\end{figure}

Consider the same Hamiltonian in the regime $h=0$; that is, the Ising model in a transverse field. Fig.~\ref{fig:tfim_crit} shows the result as we approach the critical point $J=g$ from the paramagnetic phase for $N=128$, measuring average viability throughout the algorithm. One observes a strong deterioration of the measured viability as the gap closes. Approaching the critical point, RRG takes increasingly more time than DMRG to run: runtimes for $J/g=0.6$ for both methods are shown in Table \ref{tab:tfim_times}, whereas, for example, at criticality DMRG takes 800 seconds and RRG takes 17000 seconds.

\begin{figure}[h]
\includegraphics[width=\columnwidth]{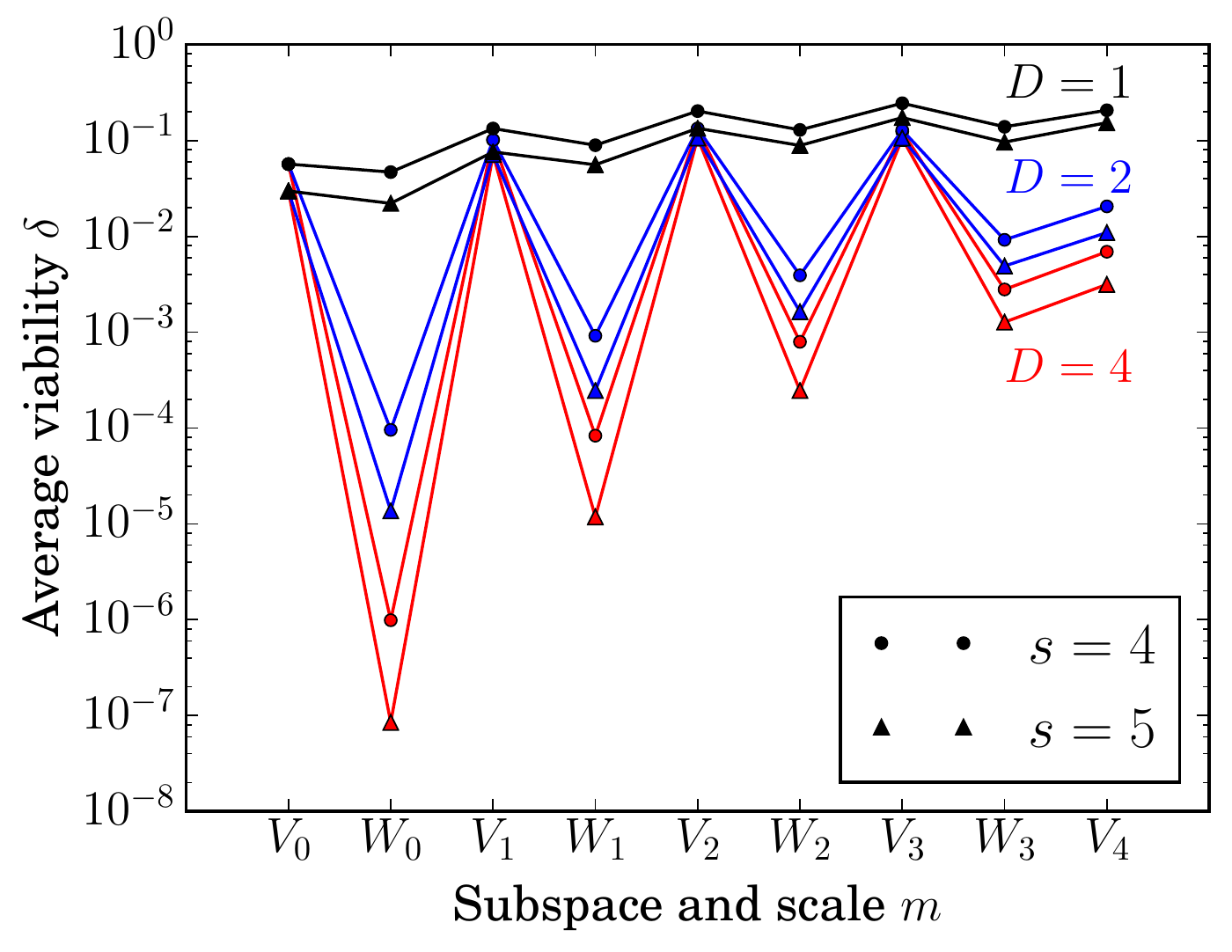}
\caption{\label{fig:tfim_prog} (color online) Viability of sets $V^\lambda_m$, $W^\lambda_m$, averaged over $\lambda$, for the transverse-field Ising model at criticality, obtained as the RRG algorithm progresses through the scale hierarchy. Data are shown for parameter values $s=4,5$ and $D=1,2,4$.}
\end{figure}

\begin{table}
\begin{tabular}{|c|c|c|}
\hline
$N$&RRG runtime (s)&DMRG runtime (s)\\
\hline
32&158&94\\
48&337&132\\
64&866&208\\
96&1871&277\\
128&3912&393\\
\hline
\end{tabular}
\caption{\label{tab:tfim_times} Runtimes of DMRG and RRG for the transverse-field Ising model with $J/g = 0.6$, using $(s,D)=(5,4)$. Some randomness is inherent in the DMRG results due to the use of random trial states. $s=5$ states are found by DMRG.}
\end{table}

We demonstrate the scaling with parameters $s$ and $D$ at criticality in Fig.~\ref{fig:tfim_prog}. The improvement in viability with increasing $D$ is less dramatic than seen in Fig.~\ref{fig:nis_prog}, corresponding to a flatter spectrum of Schmidt values across the cuts between subsystems. Note in this case that at the critical point, as the algorithm progresses, the average viability of the $V_m$ sets visibly approaches unity, in contrast to the gapped case, which appears to maintain viability bounded away from 1.

\subsection{Bravyi-Gosset model}
\label{sec:n3}
This model was initially introduced as a classification scheme for frustration-free 2-local Hamiltonians.\cite{bravyi15} The Hamiltonian is
\begin{equation}
H = \sum_{i=0}^{N-2} \ket \psi \bra \psi_{i,i+1} ,
\label{eq:bg}
\end{equation}
where $\ket \psi$ is a generic state on two qubits. Up to a global phase, such a state can be specified in the form $\ket \psi = R(\theta)_1 \left(p\ket{00} + \sqrt{1-p^2}\ket{11}\right)$, with $R(\theta)_1$ a rotation performed on the first qubit. As the spectrum is invariant under global rotation, the Hamiltonian is fully specified by the two parameters \mbox{$\theta \in [0,2\pi)$}, \mbox{$p \in [0,1/2]$}. Restricting to $\theta=0$, we may rewrite Eq.~\eqref{eq:bg} in a more familiar notation:
\begin{align}
H =& \sum_{i=0}^{N-2} \left( \frac{\sqrt{p(1-p)}}{2} \left(\sigma^x_i \sigma^x_{i+1} - \sigma^y_i \sigma^y_{i+1}\right) + \frac{1}{4} \sigma^z_i \sigma^z_{i+1}\right)\nonumber\\
&\quad+ \sum_{i=0}^{N-1} \left( \frac{1-2p}{4}\sigma^z_i + \frac{1}{8}\right)
\end{align}
That is, this model is equivalent to a particular XYZ model in a fine-tuned field. For any value of $p$ the system exhibits $(N+1)$-fold ground state degeneracy. Basis states for the ground space can roughly be thought of as having two regions of differing magnetization, with an interface which can be located at any site with ground state energy $\eps_0 = 0$. (Refer to \textcite{bravyi15} for a full description.) Therefore the algorithm choice $s \geq N+1$ is sufficient to obtain the full ground space.

\begin{figure}[ht]
\includegraphics[width=\columnwidth]{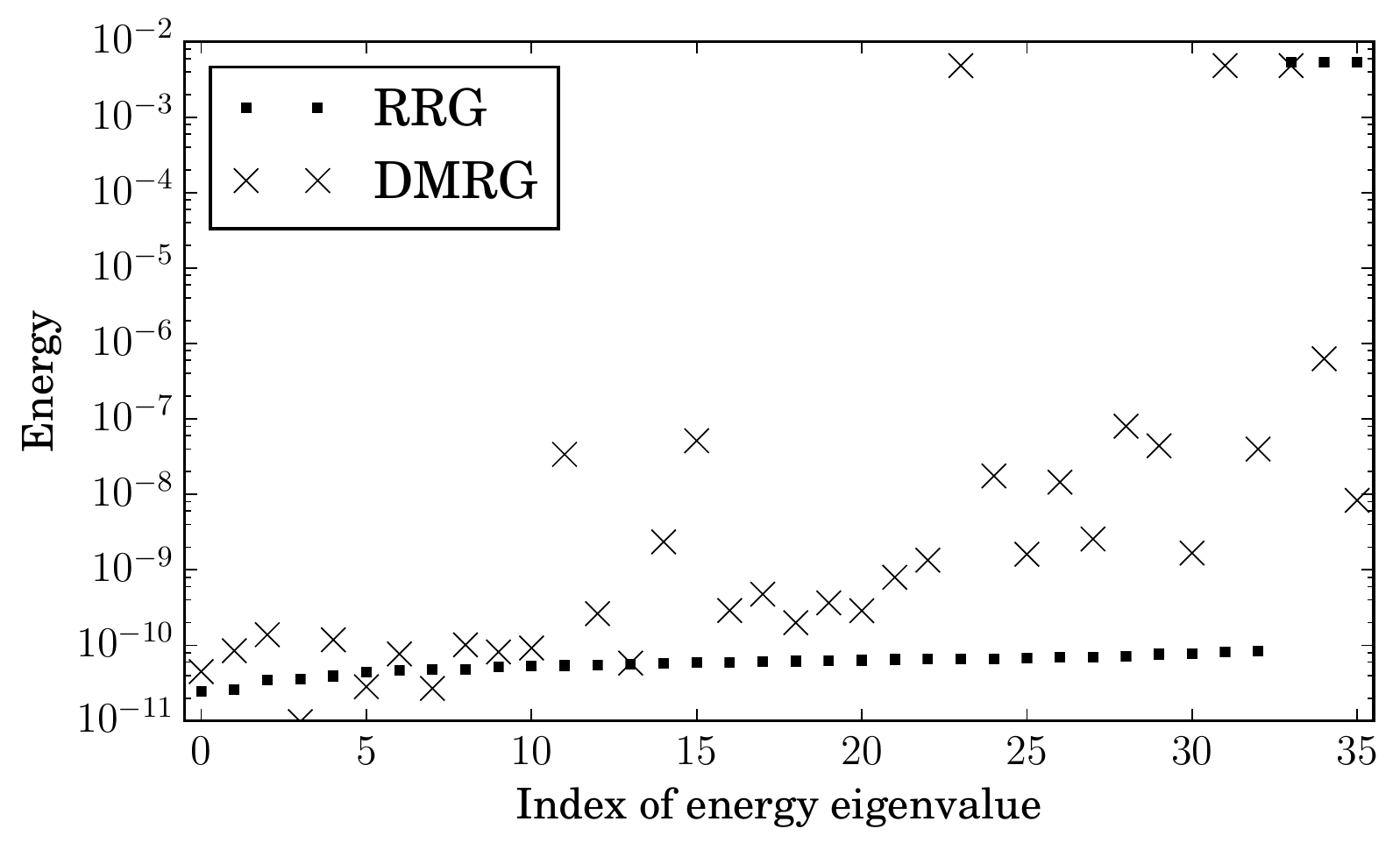}
\caption{\label{fig:gosset_spec}Energy eigenvalues of Bravyi-Gosset model with $N=32$ sites within the subspace obtained by RRG for $(s,D)=(36,3)$. Also shown are DMRG results for the 36 lowest-energy states.}
\end{figure}

The low-energy spectrum obtained by RRG for this model at $N=32$ is shown in Fig.~\ref{fig:gosset_spec}, along with the DMRG results. We use $p=1/2$; that is, $\ket \psi$ is a Bell state. Using $(s,D) = (36,3)$, RRG identifies the full zero-energy ground space to within an accuracy determined by $\tau$, the truncation error of the MPS. In contrast, obtaining the full ground space of this model is challenging for DMRG, which becomes hampered by candidate states of very high entanglement, often requiring a bond dimension an order of magnitude larger than those of RRG candidate states in order to achieve similar truncation error. These not only are computationally intensive to optimize, but also present DMRG with difficulty finding further excited states, as the modified Hamiltonian includes nonlocal projectors. Thus, the candidate states are not accurate eigenstates of the original Hamiltonian. This difficulty is evident in run times as well; to obtain the data shown took 10 hours for RRG and 40 hours for DMRG. Here we use DMRG without taking into account the degenerate ground state manifold, and we consider these results to be only a point of reference. Use of a specialized approach like multiple targeting could improve accuracy, or diagonalization of the original Hamiltonian within the subspace spanned by the DMRG candidate states could recover much of the ground space; however, no such specialized approach is needed for RRG.

\subsection{Random XY model}
\label{sec:n4}

The random XY model is an inhomogeneous spin-1/2 system with Hamiltonian
\begin{equation}
H = \sum_{i=0}^{N-2} J_i (\sigma^x_i \sigma^x_{i+1} + \sigma^y_i \sigma^y_{i+1}) ,
\end{equation}
where the position-dependent coupling constants $J_i$ are drawn from a random distribution. If the logarithm of the distribution is broad, Dasgupta-Ma real-space renormalization group analysis identifies the ground state as the random singlet phase, in which pairs of spins form singlet states at all length scales.\cite{bhatt82,ma79,dasgupta80,fisher94} This model is tractable by the Jordan-Wigner transformation, which maps onto free spinless fermions. We use this system as a benchmark of algorithmic ability to encode long-range correlations in the ground state.

\begin{figure}[h]
\includegraphics[width=\columnwidth]{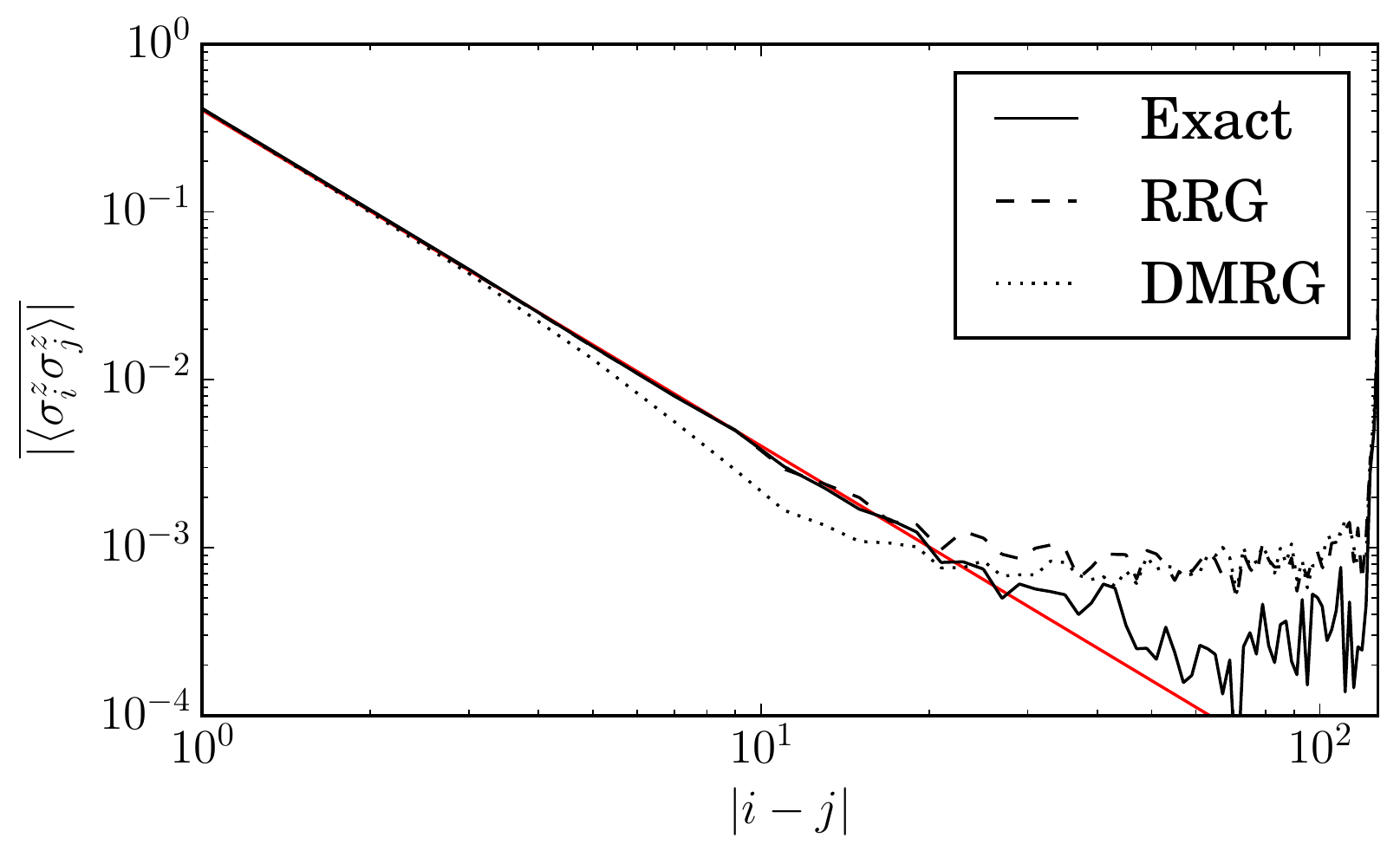}
\caption{\label{fig:rxy_szsz}(color online) Disorder-averaged decay of correlations of candidate ground states of the random XY model for $N=128$, as compared to exact results obtained through the Jordan-Wigner transformation. The predicted power-law behavior is indicated by the red line.}
\end{figure}

\begin{figure}[h]
\includegraphics[width=\columnwidth]{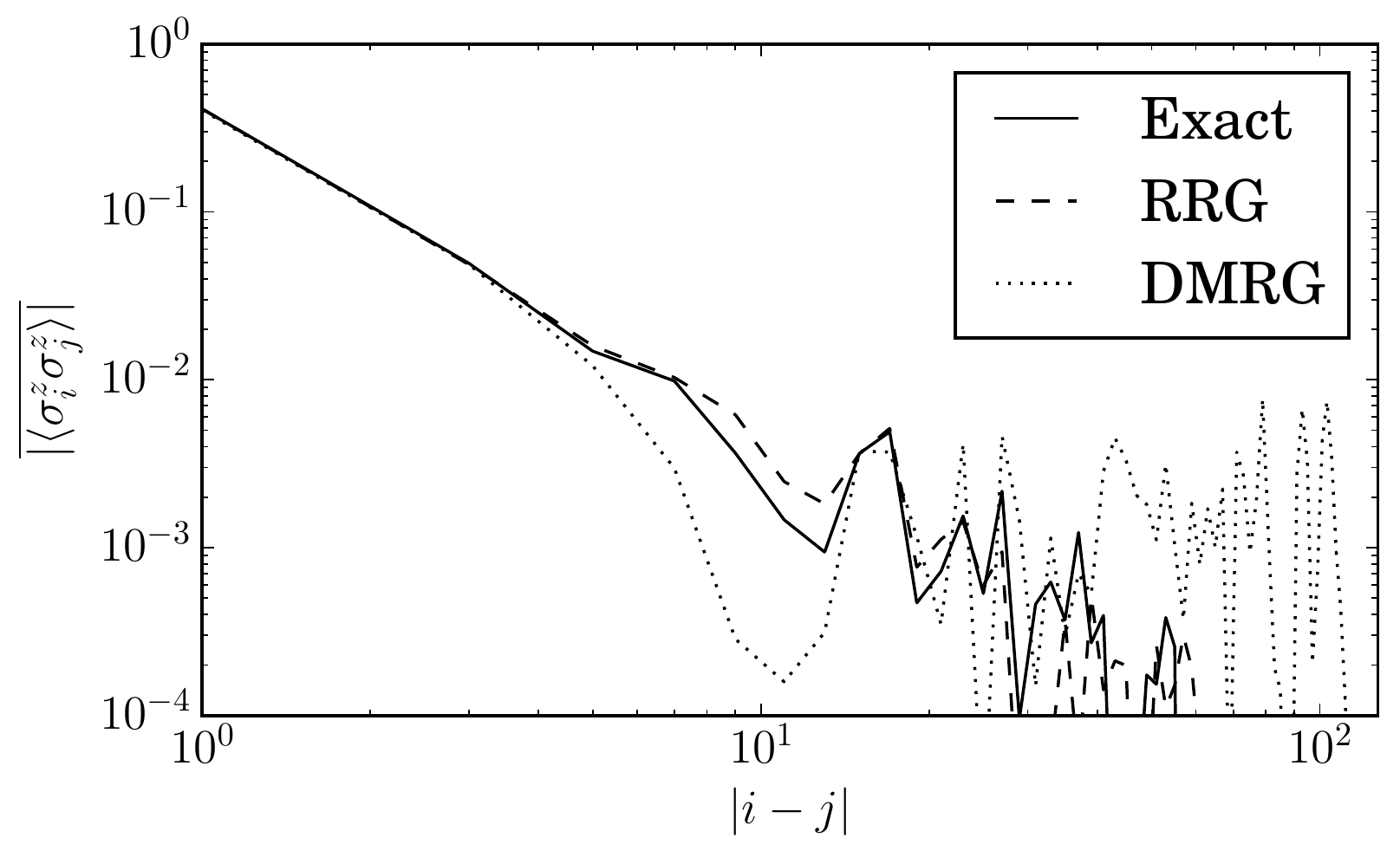}
\caption{\label{fig:rxy_ex1}A typical ``hard'' instance contained in the disorder average above, with energy gap $\gamma \approx 10^{-7}$. This is sufficiently large for RRG to track the long-range correlations with $(s,D)= (4,5)$. DMRG displays a tendency for lower correlations until saturating at the noise floor.}
\end{figure}

We use the following distribution for the Hamiltonian terms: $p(J_i) = \frac{1}{\Gamma} {J_i}^{-(1-\frac{1}{\Gamma})}$, $J_i \in (0,1]$, with $\Gamma$ controlling the width of the distribution of log-energies.\cite{fisher94} We fix $\Gamma = 2$, which is sufficiently broad that the ground state is composed predominantly of localized singlet states on neighboring sites, along with spatially separated correlated qubits occurring at all length scales. As a metric we use the average two-point correlation function $\overline{\langle \sigma^z_i \sigma^z_j \rangle}$ as a function of separation $r=|i-j|$ in the ground state, which is known to decay algebraically as $r^{-2}$. This quantity is compared to exact diagonalization results from the inhomogeneous free fermion description in Fig.~\ref{fig:rxy_szsz}.

\begin{figure}[h]
\includegraphics[width=0.79\columnwidth]{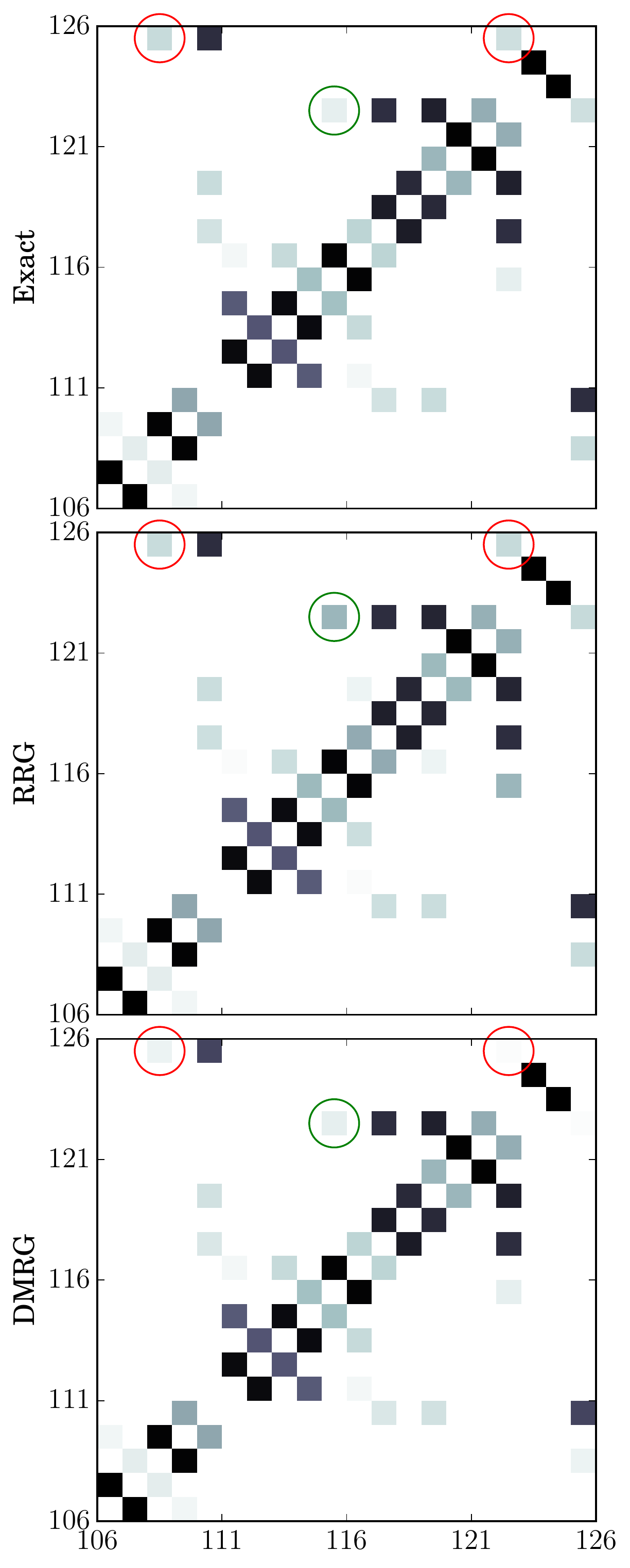}
\caption{\label{fig:rxy_ex2} (color online) Expectation value $\langle \sigma^z_i \sigma^z_j \rangle$, where sites $i$ and $j$ are given by the axes for $i,j \in [106,125]$. The color scales with $\log |\langle \sigma^z_i \sigma^z_j \rangle|$ and runs from $[-2,0]$ in all plots, with darker color indicating a higher value. The diagonal is omitted. Circles mark particular sites where differences between exact results and candidate states are evident. This disorder realization is the same ``hard'' instance shown in Fig.~\ref{fig:rxy_ex1}.}
\end{figure}

These results are intended to present a fair comparison between DMRG and RRG. Both methods used unrestricted MPS bond dimension to achieve a truncation error $\tau \leq 10^{-12}$. Typically the ground state bond dimension is similar for both methods. The RRG parameters are $(s,D) = (4,5)$. DMRG used 20 sweeps per state, and convergence of several ``hard'' examples (see below) was confirmed using 50 sweeps. DMRG typically took 1 hour to converge $s=4$ states and RRG took 8 hours to complete. The average is over 150 disorder realizations.

The observed ``saturation'' of the correlations of Fig.~\ref{fig:rxy_szsz} to a noise floor arises from the structure of the low-energy excited states. For a broad initial distribution, the energy gap of a specific disorder Hamiltonian may be very small. For any method using MPS, a lower limit on the gap in order to distinguish the ground state (at energy $\eps_0$) is $\gamma \sim \tau \eps_0$, below which the MPS truncation procedure will randomly select a vector from the low-lying subspace. However, even for realizations with much larger gaps a candidate ground state may include substantial contributions from low-energy excited states. A singlet of length $l$ has energy scale $\eps \sim e^{-\sqrt l}$; thus, the low-lying states involve excitations localized on the long-range entangled sites. Choosing a random superposition of these amounts to white noise at long distances. Instances of such Hamiltonians in the disorder average must necessarily eventually overwhelm the decay of correlations; here the distribution of energy gaps is very broad on a log scale,\cite{fisher98} so these cases are frequent. However in all cases the RRG candidate state has $\O(1)$ overlap with the true ground state, and typically this overlap is greater than 99\%.

For disorder-averaged correlations at short range up to $|i-j| \approx 20$, RRG reproduces algebraic decay of correlations matching the exact results. In contrast, the DMRG candidate states demonstrate stronger decay of correlations. There is no systematic difference in MPS bond dimension between DMRG and RRG, indicating that RRG is not simply using additional resources, but is indeed more sensitive to long-range correlations.

Independent of the saturation due to the energy gap, the disorder average comprises both ``easy'' and ``hard'' instances. In easy cases both DMRG and RRG match the exact results closely at all length scales. In the hard cases both algorithms obtain the correlations only approximately, but DMRG appears to consistently underestimate correlations. RRG does not demonstrate a tendency towards either enhanced or reduced correlations. We provide an example of the spatially averaged correlations from a hard disorder realization in Fig.~\ref{fig:rxy_ex1}. Fig.~\ref{fig:rxy_ex2} shows an example of measured correlations $\langle \sigma^z_i \sigma^z_j\rangle$ for various sites $i,j \in [106,125]$ in this particular disorder realization. Each square corresponds to a measurement $\langle \sigma^z_i \sigma^z_j \rangle$ where $(i,j)$ are specified by the axes. Darker squares indicate a larger magnitude of correlation between these sites. We show the exact results, RRG, and DMRG, and indicate some particular pairs of sites where either DMRG (red) or RRG (green) differ visibly from exact results. These variations in certain entangled sites tend toward reduced correlations in DMRG candidate ground states; it is unclear how much additional sweeping is required to compensate. RRG shows similar inaccuracies, but these are random, due to states missing from certain viable sets. Accurate correlations emerge in the disorder-averaged value, and the performance on individual disorder realizations can be controllably improved by tuning the dimension of the viable sets through the parameters $s$ and $D$.  

\section{Discussion}
\label{sec:discussion} 

DMRG has long been the method of choice for numerical calculations involving ground states of 1D systems, and over time both its efficiency and range of applicability have gone through multiple improvements and extensions. One of the main findings of our initial numerical investigation is that the RRG algorithm, developed for theoretical purposes, can in fact be made quite effective in practice, to the point of providing a potentially viable alternative to DMRG in certain cases of practical interest. We stress that the choices of parameters that we employ in our numerics are quite far from the theoretically guaranteed regime. Additionally, many of the building blocks required for the proof have been altered in our implementation. Therefore the strict guarantees no longer hold. Regardless, we find that RRG obtains ground state candidates having large overlap with the true ground state in a variety of physically relevant models, and surpasses existing techniques in obtaining low-energy excited states and ground states of particular models demonstrating large degeneracy or long-range entanglement.

Like another numerical scheme, {\it time-evolving block decimation} (TEBD), the RRG procedure is a projector method, relying on operators extracted from the AGSP to guide the choice of states between scales. As a result, given a sufficiently accurate AGSP, RRG will not output a part of the spectrum strictly excited above the ground space. This is advantageous relative to variational ansatz methods which may without warning converge to an excited state rather than the ground state. (For example, if the energy landscape in Hilbert space has local minima or is very flat in the low-energy space, as is the case with the random XY model of Sec.~\ref{sec:n4}.) The downsides to projector methods are that performance strongly depends on the gap and that a random initial state, even taken from the manifold of low bond dimension MPS, has exponentially small overlap with the ground state. RRG circumvents the latter issue by never choosing a trial wavefunction on the entire system, but rather building global states from wavefunctions supported on blocks which {\it already have good viability}; thus the projection step never has to overcome starting with an exponentially small overlap between the initial and the target state.

At present the run times required by the algorithm remain a challenge. Thus, the feasibility of RRG as a numerical method is essentially determined by the scaling discussed previously. This situation does invite future improvements. Some are immediate: for example, one may exploit symmetries of a particular problem (say, reflection symmetry across the middle of the system) in order to reduce duplication of work. Other improvements to the current implementation are more technical. For example, as described here the management of subspaces is clumsy: operations such as  addition of MPS necessitate keeping careful track of gauge and add computational overhead for what is in principle a simple procedure. The use of data structures more appropriate to these operations could ameliorate scaling problems in all steps of the algorithm.\footnote{Since submission of this manuscript, the authors have significantly advanced the state of the RRG code, which may be found online at \texttt{https://github.com/brendenroberts/RigorousRG}}

Indeed, an advantage of RRG is precisely this flexibility, to operate independently of a specific representation of states in Hilbert space. Here we have described an MPS RRG. In order to translate the logic to subspaces whose basis states are described by MERA---as would be natural for critical phases---one needs only the ability to perform evaluation of observables and addition. The former is a standard contraction which is highly efficient in MERA, and the latter can be seen as a variational process on overlaps, providing a straightforward interpretation as a MERA operation. Systems with periodic boundary conditions also present an interesting generalization, as until the final level the steps of the algorithm are insensitive to the system boundaries, provided an appropriate AGSP is given. On a more speculative note, other tensor network ansatze may also be amenable: although it is not known that the RRG algorithm scales efficiently in higher dimensions, the hierarchical structure does generalize in an evident way and it may be the case that the algorithm gives acceptable results for PEPS representations of some two-dimensional systems. 

Our numerical results suggest situations in which RRG may perform well relative to existing algorithms. The first, informed by Sec.~\ref{sec:n1}, is a case in which localized and delocalized excitations lie close in energy. An optimization algorithm operating on local degrees of freedom in a sweeping pattern may exhibit a bias towards delocalized excitations, which allow for effective optimization on many lattice sites. RRG is largely insensitive to such distinctions. The second case is that of Sec.~\ref{sec:n3}, exhibiting highly degenerate ground states. The full ground space is more accurately found in its entirety by RRG than DMRG. The iterative DMRG procedure of finding states is susceptible to finding poor or highly entangled candidates, which reduce the accuracy of subsequent candidates. Such a limitation is not fundamental and could likely be eliminated by modification of the procedure; however no such modification is necessary for RRG. Finally, in Sec.~\ref{sec:n4} we observe in the random XY model in the random singlet phase that long-range correlations are encoded more precisely in the ground state candidate of RRG than of DMRG, influencing observables computed for the state.

The examples we provide illustrate specific properties indicating that a model may be well suited for RRG. However, very little is known about its more general performance: other systems with disorder, periodic boundary conditions, and higher dimensions all pose interesting challenges and could constitute exciting new directions within this formalism.

\begin{acknowledgements}
We acknowledge useful discussions with M.~Fishman and S.~White's research group, as well as with C.~White and C.-J.~Lin. The numerical results were computed with the ITensor library\cite{itensor} of E.~Stoudenmire and S.~White. This work was supported by the Institute for Quantum Information and Matter, an NSF Physics Frontiers Center, with support of the Gordon and Betty Moore Foundation. Additional funding support was provided by the NSF through Grant DMR-1619696.
\end{acknowledgements}

\bibliography{refs.bib}

\appendix

\section{Differences from~\textcite{arad17}}
\label{sec:cspaper} 

In this appendix we give a detailed account of the main points of departure of our numerical procedure from the theoretically guaranteed algorithm introduced in~\textcite{arad17}, giving heuristic justification for our choices. We refer to the paper for a more thorough introduction to the main concepts discussed here, such as the notion of viable set and AGSP.
 
For concreteness we base our comparison on the algorithm presented in~\textcite{arad17} for the case of a local Hamiltonian with degenerate gapped ground space (Assumption (DG)). The algorithm is stated as Algorithm~1 in~\textcite{arad17}. It consists of two main steps, \emph{Generate} and \emph{Merge}. The two steps together recursively construct a sequence of viable sets $V_m^\lambda$ for an $N$-qubit local Hamiltonian, where as in the main text $m$ denotes a scale parameter and $\lambda$ indexes a subregion. 

\subsection{Generate}
\label{sec:cs-generate}

The goal of the \emph{Generate} step is to generate an MPO representation for a suitable AGSP. In~\textcite{arad17} a fresh AGSP is computed for each scale $m$ and region $\lambda$. Given a decomposition $\H = \H_L\otimes \H_m^\lambda \otimes \H_R$, a global AGSP is defined as $K_m^\lambda = T_k(\tilde{H})$, where $\tilde{H}$ is a norm-reduced approximation of $H$ (which depends on the region decomposition) and $T_k$ a suitably scaled Chebyshev polynomial of degree $k$. The operators $A_{m,r}^\lambda$ are then computed from a specific  decomposition of $K_m^\lambda$ across the left and right boundaries, yielding $D^2$ terms $A_{m,r}^\lambda$ such that the expansion procedure  $V^\lambda_m \to W^\lambda_m $ described in the main text is guaranteed to have a significant improvement on the viability parameter.  

Here we depart from the theoretical algorithm in two important ways. First we use a simpler construction of AGSP, which we expect to exhibit similar behavior but is more efficient to compute. Our AGSP takes the form of an approximation $K\approx e^{-kH/t}$ obtained by Trotter decomposition. (In~\textcite{arad17} a similar approach is taken to norm-reduce the parts of the Hamiltonian that lie in the regions $L,M$ and $R$ but are a distance at least $\ell>0$ from the boundaries.) In~\textcite{arad17} the properties of the Chebyshev polynomial are essential to establish that the AGSP has sufficiently low bond dimension across the boundaries of region $M$. Considering only the efficiency in terms of improvement in viability, however, the use of $e^{-kH/t}$ over the whole chain gives similar guarantees.  

Using our simpler construction implies a loss of theoretical control over the bond dimension $D$ of the AGSP operator across the left and right cuts. This entails a second main point of departure from the theoretical algorithm, as a choice has to be made as to which operators $A_{m,r}^\lambda$ to keep. As described in the main text we proceed in a natural way by considering the MPO as an MPS and performing SVD operations to create virtual bonds between sites. We then make the choice of keeping operators associated with the $D^2$ highest Schmidt weights. This choice is heuristic: the Schmidt weights control the Frobenius norm of the associated term $A_{m,r}^\lambda$, rather than the operator norm of the resulting operator, as would be desirable. The heuristic nevertheless proved effective: in practice the magnitude of the Schmidt coefficients often fell off quickly, allowing for a relatively aggressive choice of cutting point.

\subsection{Merge process}
\label{sec:cs-merge}

The second step in the algorithm is called \emph{Merge}. The goal of this step is to combine two neighboring viable sets into a single viable set over the union of the two regions, with similar approximation and size guarantees. The procedure is described as \emph{Merge'} in~\textcite{arad17}. Merge' is provided as input viable sets $W_m^{\lambda}$ and $W_m^{\lambda+1}$ defined over neighboring regions, and returns a viable set $W_{m+1}^{\lambda/2}$ defined over the union of the two regions. Merge consists of three steps: \emph{Tensoring}, \emph{Random Sampling}, and \emph{Error Reduction}. 

\begin{enumerate}
\item \emph{Tensoring:} This step is the same as in~\textcite{arad17}. 

\item \emph{Random Sampling:} Here as already mentioned in the main text we depart from~\textcite{arad17} in an important way. In~\textcite{arad17} a family of $s$ vectors lying in $W_m^{\lambda}\otimes W_m^{\lambda+1}$ is obtained by random sampling within the subspace. In practice this procedure is very inefficient: (i) it requires performing high-weight (random) linear combinations of MPS, a step that is computationally expensive due to the MPS renormalization procedure; (ii) the linear combinations formed tend to be arbitrary, and in particular their MPS representations may have high MPS bond dimension, as each vector may include an ``irrelevant'' (with respect to the low-energy eigenspace of the Hamiltonian) component that artificially inflates its complexity. 

Here we replace random sampling by a deterministic choice of the $s$ lowest-energy eigenvectors of the restriction of $H$ to $W_m^{\lambda}\otimes W_m^{\lambda+1}$. The idea is that low-energy eigenstates are likely, due to the local structure of the Hamiltonian, to display less entanglement. Indeed in practice this procedure is much more efficient, and yields MPS with lower bond dimension, than the random sampling proposed in~\textcite{arad17}. 

However, there is a priori no reason for the low-energy eigenstates of the block Hamiltonian to form a viable set for the global low-energy space. A simple heuristic argument can nevertheless be given to argue correctness of our procedure. Recall that the viability criterion Eq.~\eqref{eq:def-viable1} guarantees that the initial tensor product space supports a good approximation to any ground state. Considering the Schmidt decomposition of this approximation, each of the Schmidt vectors will have a certain energy with respect to the block Hamiltonian $H_{m+1}^{\lambda/2}$, which may not be minimal. The key is thus to argue that vectors with high energy will not have an important contribution to the Schmidt decomposition of the ground state. In general approximation error and energy difference can scale with the norm of the Hamiltonian, making the argument difficult. However, for the purposes of approximating the ground space of a local Hamiltonian two elements play in our favor: first, locality of $H$, and second, the area law. The former allows to show that the low-energy space of $H$ is well-approximated by an approximation of $H$ with constant norm, so that the error blow-up mentioned above can be controlled (see~\textcite[Proposition 3]{arad17}, for a precise statement). The latter establishes that the ground state has low bond dimension, so that few Schmidt vectors need to be considered (see~\textcite[Lemma~15]{arad17}, for details on how this can be used). Together these two properties provide a heuristic argument in favor of our modified procedure.

\item\emph{Error Reduction:} The goal of this step is to improve the approximation quality of the viable set. We follow the procedure described in~\textcite{arad17}, except that the operators $\{A_{m,r}^\lambda\}$ are generated differently, as already described. \end{enumerate}

The final iteration is performed on two viable sets $V_{m^*-1}^0$ and $V_{m^*-1}^1$, each with support on one half of the system. The algorithm returns the low-lying energies and eigenstates obtained via exact diagonalization of the Hamiltonian restricted to the final viable subspace.
\end{document}